\documentclass[useAMS, usenatbib]{mn2e}
\usepackage{lscape}
\usepackage{rotating}
\usepackage{amssymb}
\usepackage[utf8]{inputenc}
\usepackage{float}
\usepackage{hyperref}
\hypersetup{colorlinks = true, linkcolor = blue, urlcolor=blue, citecolor=blue} 
\usepackage[normalem]{ulem}
\newcommand{\GG}[1]{}
\usepackage{tabularx}
\usepackage{rotating, booktabs}

\def\msun{\hbox{M$_\odot$}}

\def\frac{\hbox{f$_{\rm mix}$}}

\def\t4{\hbox{t$_{\rm 4}$}}

\def\cm3{\hbox{cm$^{-3}$}}

\usepackage{graphicx}
\def\Dunb{$\Delta$($Un-B$)}
\def\drgb{$\delta^{\rm RGB}$(CUnBI) \,}

\title[Age as a Major Factor in the Onset of MPs]
{Age as a Major Factor in the Onset of Multiple Populations in Stellar Clusters}
\author[Martocchia et al.] {S. Martocchia$^{1}$, I. Cabrera-Ziri$^{2}$\thanks{Hubble Fellow.}, C. Lardo$^{1,3}$, E. Dalessandro$^{4}$,  N. Bastian$^{1}$, 
\newauthor V. Kozhurina-Platais$^{5}$,  C. Usher$^{1}$,  F. Niederhofer$^{6}$, M. Cordero$^{7}$,  D. Geisler$^{8}$, 
\newauthor K. Hollyhead$^{9}$, N. Kacharov$^{10}$, S. Larsen$^{11}$, C. Li$^{12}$, D. Mackey$^{13}$, 
\newauthor  M. Hilker$^{14}$,  A. Mucciarelli$^{15}$, I. Platais$^{16}$, M. Salaris$^{1}$.\\
$^{1}$Astrophysics Research Institute, Liverpool John Moores University, 146 Brownlow Hill, Liverpool L3 5RF, UK\\
$^{2}$Harvard-Smithsonian Center for Astrophysics, 60 Garden Street, Cambridge, MA 02138, USA\\
$^{3}$Laboratoire d’astrophysique, École Polytechnique Fédérale de Lausanne (EPFL), Observatoire, 1290, Versoix, Switzerland\\
$^{4}$INAF, Osservatorio Astronomico di Bologna, via Ranzani 1, 40127, Bologna, Italy\\
$^{5}$Space Telescope Science Institute, 3700 San Martin Drive, Baltimore, MD 21218, USA\\
$^{6}$Leibniz-Institut f\"ur Astrophysik Potsdam, An der Sternwarte 16, Potsdam 14482, Germany\\
$^{7}$Astronomisches Rechen-Institut, Zentrum f\"ur Astronomie der Universit\"at Heidelberg, M\"onchhofstrasse 12-14, D-69120 Heidelberg, Germany\\
$^{8}$Departamento de Astronomia, Universidad de Concepcion, Casilla 160-C, Chile\\
$^{9}$Department of Astronomy,Oscar Klein Centre,Stockholm University, AlbaNova, Stockholm SE-10691,Sweden\\
$^{10}$Max-Planck-Institut f\"ur Astronomie, K\"onigstuhl 17, D-69117 Heidelberg, Germany\\
$^{11}$Department of Astrophysics/IMAPP, Radboud University, P.O. Box 9010, 6500 GL Nijmegen, The Netherlands\\
$^{12}$Department of Physics and Astronomy, Macquarie University, Sydney, NSW 2109, Australia\\
$^{13}$Research School of Astronomy and Astrophysics, Australian National University, Canberra, ACT 2611, Australia\\
$^{14}$European Southern Observatory, Karl-Schwarzschild-Stra\ss e 2, D-85748 Garching bei M\"unchen, Germany\\
$^{15}$Department of Physics and Astronomy, University of Bologna, Viale Berti Pichat 6/2, I-40127 Bologna, Italy\\
$^{16}$Department of Physics and Astronomy, Johns Hopkins University, 3400 North Charles Street, Baltimore, MD 21218, USA\\
}
\date{Accepted. Received ; in original form.}
\pagerange{\pageref{firstpage}--\pageref{lastpage}}
\pubyear{2017}

\begin{document}

\maketitle
\label{firstpage}

\begin{abstract}	
 It is now well established that globular clusters (GCs) exhibit star-to-star light-element abundance variations (known as multiple stellar populations, MPs). Such chemical anomalies have been found in (nearly) all the ancient GCs (more than 10 Gyr old) of our Galaxy and its close companions, but so far no model for the origin of MPs is able to reproduce all the relevant observations. To gain new insights into this phenomenon, we have undertaken a photometric Hubble Space Telescope survey to study clusters with masses comparable to that of old GCs, where MPs have been identified, but with significantly younger ages. Nine clusters in the Magellanic Clouds with ages between $\sim$ 1.5-11 Gyr have been targeted in this survey. We confirm the presence of multiple populations in all clusters older than 6 Gyr and we add NGC 1978 to the group of clusters for which MPs have been identified. With an age of $\sim$ 2 Gyr, NGC 1978 is the youngest cluster known to host chemical abundance spreads found to date. We do not detect evident star-to-star variations for slightly younger massive clusters ($\sim$ 1.7 Gyr), thus pointing towards an unexpected age dependence for the onset of multiple populations. This discovery suggests that the formation of MPs is not restricted to the early Universe and that GCs and young massive clusters share common formation and evolutionary processes.
\end{abstract}


\section{Introduction}
\label{sec:intro}

Globular clusters (GCs) were traditionally thought to be composed of simple stellar populations (SSPs), with all the stars in a given cluster having the same age and initial elemental abundances, within some small tolerance. 
However, observational evidence collected in the last decade has conclusively demonstrated that GCs host multiple populations (MPs) of stars, which manifest in the form of distinctive star-to-star light element variations. While some GC stars have the same C, N, O, Na abundances as field stars with the same metallicity (i.e., iron content), a significant fraction of cluster members show systematically enhanced N and Na along with depleted C and O \citep{cannon98,carretta09}. Almost all old ($>$ 10 Gyr) and massive (above a few times $10^4$ \msun) GCs in the Milky Way \citep{piotto15} and nearby galaxies \citep{mucciarelli09,dalessandro16,paperI,larsen12} have been found to host star-to-star chemical anomalies. 

Multiple models for the origin of MPs have been put forward. 
In the most popular MP formation models, a second generation of stars is formed from the ejecta of massive stars from a first generation, diluted with some amount of unprocessed material with pristine composition. While the stellar sources for enrichment are different in different scenarios \citep{decressin07,dercole08,demink09}, none of the proposed models is able to reproduce the main observational properties of MPs without making ad hoc assumptions \citep{bastian15}. We still lack a self-consistent explanation of the physical processes responsible for the MPs phenomenon as well as an understanding of which (if any) cluster properties controls whether a GC will host chemical anomalies or not. 

While MPs are commonly detected in massive and old clusters, light elements variations are not found in clusters of comparable age but lower mass (e.g., E3, which has a mass of $10^4$ \msun , \citealt{salinas15}). Such evidence led many authors to suspect cluster mass as the key parameter controlling whether a GC will host MPs \citep{gratton12}. Nonetheless, the lack of chemical spreads in relatively young ($\sim 1-2$ Gyr, \citealt{mucciarelli14}), but still massive (few times $10^5$ \msun) clusters suggests a potentially more complex cause. It has even been proposed that the ancient GCs are intrinsically different from the young clusters with similar masses, as only ancient GCs have been found to host MPs \citep{carretta10}.

It is important to stress here that {\it chemical MPs} detected in clusters older than 6 Gyr are intrinsically different from the multiple populations associated with the extended main sequence turn-offs (eMSTO) detected in the large majority of clusters younger than 2 Gyr (e.g. \citealt{milone09}). In the latter, no star-to-star abundance variations in light elements have been found (e.g. \citealt{mucciarelli08,mucciarelli14}). Instead, the spread in colours in the turn-off region observed in their colour-magnitude diagrams (CMDs) may be likely due to stellar rotation \citep{bastian16,bastian17}.

We present here results from our Hubble Space Telescope (HST) photometric survey. This consists in HST UV observations of 9 star clusters in the Magellanic Clouds (MCs). 
All clusters have similar masses ($\gtrsim 10^5$ \msun) but span a wide range of ages (from 1.5 up to 11 Gyr). 
The main goal of our survey is to test whether MPs are exclusively found in ancient GCs, and hence, to shed light on the physical property that controls the onset of MPs.

To date, our survey has demonstrated that MPs are present in all clusters older than 6 Gyr \citep{paperI,paperII}; however, they were not detected in a $\sim$ 1.5 Gyr old cluster (NGC 419, \citealt{paperIII}).
We report here the final results from our survey by adding the four remaining clusters, namely NGC 1978, NGC 1783, NGC 1806, and NGC 1846, to our analysis.

We will show that multiple populations are present in the red giant branch (RGB) of the $\sim$ 2 Gyr Large Magellanic Cloud (LMC) cluster NGC 1978. This is the youngest cluster found to host chemical anomalies so far. 
On the contrary, we report no detection of MPs for the younger LMC clusters, the $\sim$ 1.7 Gyr old NGC 1783, 1806 and 1846.

This paper is organised as follows: in \S \ref{sec:obs} we will report on the photometric reduction procedures for NGC 1978, while we outline the analysis used to quantify the detection of MPs in \S \ref{sec:analysis}. In \S \ref{sec:results} we present the results from our HST survey. Section \S \ref{sec:models} provides a description of our stellar atmospheres models and their comparison with NGC 1978 data. We discuss our results in \S \ref{sec:discussion}. We will present the detailed analysis of the younger clusters NGC 1783, 1806 and 1846 in a follow-up paper.

\begin{figure}
	\centering
	\includegraphics[scale=0.38]{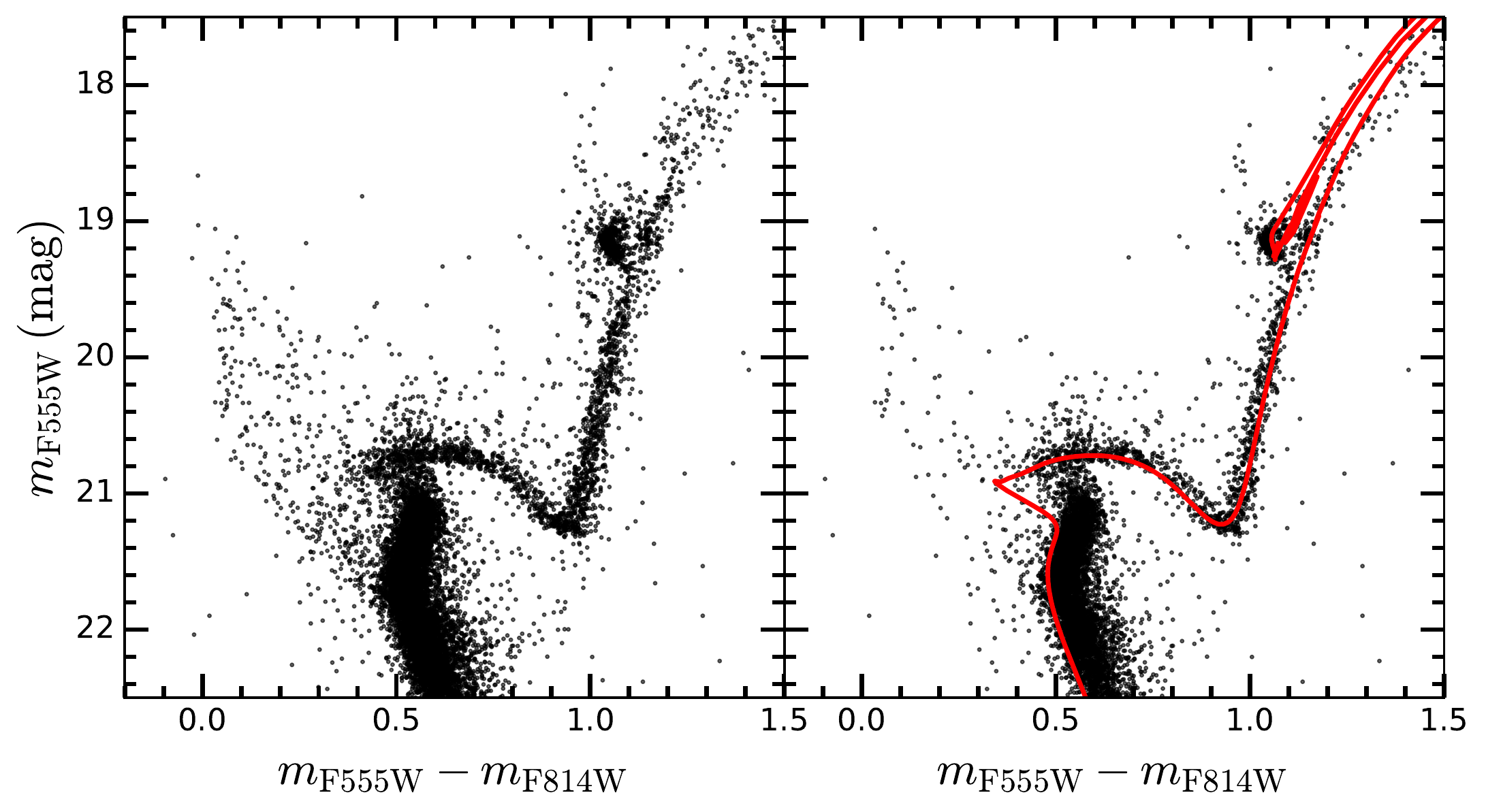}	
	\caption{$V-I$ vs. $V$ CMD of NGC 1978 before (left panel) and after (right panel) the field star subtraction. The red curve in the right panel represents the MIST isochrone we adopted for NGC 1978. The parameters used to derive it are: $\log$(t/yr) $ = 9.35$ (corresponding to $\sim$ 2.2 Gyr); distance modulus $M-m = 18.5$; $E(B-V) = 0.07$; metallicity [Fe/H] $= -0.5$.}
	\label{fig:S1}
\end{figure}

\begin{figure*}
	\centering
	\includegraphics[scale=0.6]{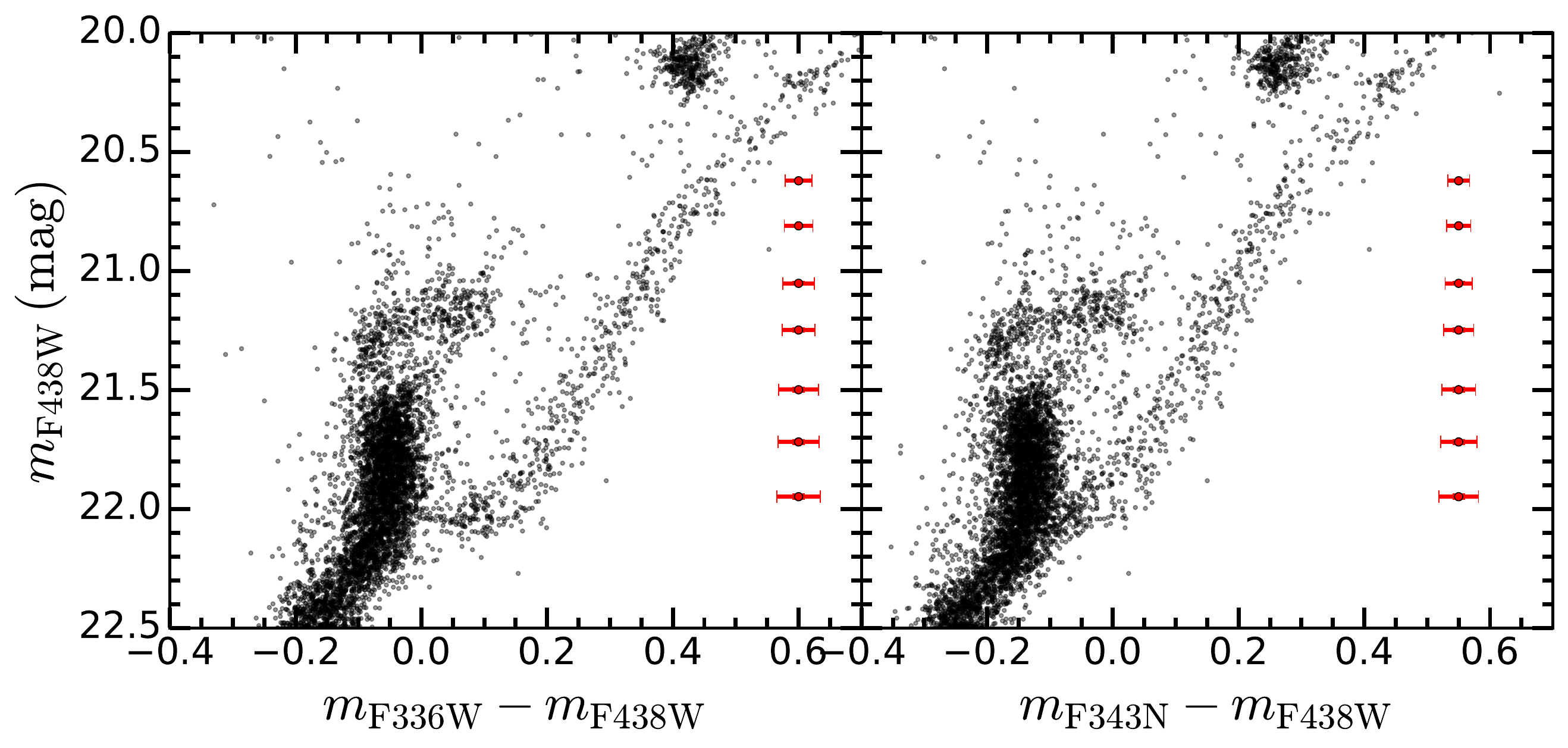}
	\caption{$U-B$ vs. $B$ (left panel) and $Un-B$ vs. $B$ (right panel) CMD of NGC 1978. On the right side, the median of photometric errors in colour and magnitude is reported as red filled circles in bins of $\sim$ 0.2 mag.}
	\label{fig:S2}
\end{figure*}

\section{Observations and Data Reduction}
\label{sec:obs}

The photometric catalogue of NGC 1978 was obtained through HST observations from our ongoing survey (GO-14069 program, PI: N. Bastian, \citealt{paperI}). 
We used three new filters, i.e. F336W, F343N and F438W (WFC3/UVIS camera), and we complemented our analysis with archival observations in F555W and F814W filters (ACS/WFC, GO-9891, PI: G. Gilmore). 

As thoroughly discussed in previous papers \citep{paperI, paperII}, the adopted UV filters (F336W, F343N and F438W) are sensitive to elements associated with MPs (mainly N, and C to a lesser degree), as they include strong NH, CN, and CH molecular absorption features within their passbands. MPs are identified by observing splits or spreads at different stellar evolutionary stages in high-precision CMDs constructed using photometry with the adopted filters \citep{sbordone11}.

The UV WFC3 dataset consists of three images for the F336W ($U$ hereafter) filter (one exposure of 380, 460 and 740 s each), 6 images for the F343N, $Un$ hereafter (one exposure of 425, 450, 500, 1000s each and two of 800 s) and another 6 images for the F438W, $B$ hereafter (one exposure of 75, 120, 420, 460, 650, 750 s each). Optical images consist of one each for the F555W and F814W filters ($V$ and $I$ hereafter), with exposures of 300 s and 200 s, respectively.

The images have been processed, flat-field corrected, and bias-subtracted by using the standard HST pipeline ($flc$ images). Pixel-area effects have been corrected by applying the Pixel Area Maps images to each image. We also corrected all images for cosmic rays contamination by using the L.A. Cosmic algorithm \citep{vandokkum01}.

The photometric analysis has been performed following the same strategy as in \cite{dalessandro14}. Briefly, we used DAOPHOTIV \citep{stetson87} independently on each camera and each chip. We selected several hundreds of bright and isolated stars in order to model the point-spread function (PSF). All available analytic functions were considered for the PSF fitting (Gauss, Moffat, Lorentz and Penny functions), leaving the PSF free to spatially vary to the first-order. In each image, we then fit all the star-like sources with the obtained PSF as detected by using a threshold of 3$\sigma$ from the local background. Next, we took advantage of ALLFRAME \citep{stetson94} to create a star list by starting with stars detected in at least three images for the $U$ band and four images in the $Un$ and $B$ bands (one image for $V$ and $I$ filters). The final star lists for each image and chip have been cross-correlated by using DAOMATCH, then the magnitude means and standard deviation were obtained through DAOMASTER. The final result consists in a catalogue for each camera. We exactly repeated each step in the above photometric analysis by using a third-order spatial variation for the PSF, as an additional check for the quality of the PSF and, hence, to strengthen our results. No significant changes were detected by comparing the photometric catalogues with different PSF spatial variation. However, we decided to perform the analysis on the catalogue where the PSF was left free to spatially vary to first-order, as PSF residuals were lower in such an analysis. 
 
Instrumental magnitudes have been converted to the VEGAMAG photometric system by using the prescriptions and zero-points reported on the dedicated HST web-pages. Instrumental coordinates were reported on the absolute image World Coordinate System by using CataXcorr\footnote{Part of a package of astronomical softwares (CataPack) developed by P. Montegriffo at INAF-OABo.}. The WFC3 catalogue was combined with the ACS by using the same CataXcorr and CataComb. 
We finally selected stars based on quality parameters provided by DAOPHOT, i.e. chi, sharpness and errors on magnitudes. Stars in the final catalogue are selected having $|sharp|<0.15$ and $chi<3$, errors on $U$ and $Un$ lower than or equal to 0.05 and errors on B $\leq$ 0.3. 

In order to select cluster members, we defined an ellipse region centred in the centre of the cluster with semimajor axis of 1200 pixels and ellipticity $\epsilon$ =0.25\footnote{We define the ellipticity as $\epsilon = \sqrt{(a^2-b^2)/a^2}$, where $a$ and $b$ represent the semimajor and semiminor axes, respectively.}. It is indeed well-known from previous studies \citep{geislerhodge80,mucciarelli07} that NGC 1978 shows a non-negligible ellipticity. Then, we defined a background reference region having the same area as the cluster region in order to statistically subtract field stars from the cluster CMD in $U-B$ vs. $B$ space. We removed the closest star in colour-magnitude space in the cluster region, for each star in the background region. Fig. \ref{fig:S1} shows the $V-I$ vs. $V$ CMD of NGC 1978 before (left panel) and after (right panel) the field star subtraction.

\begin{figure*}
	\centering
	\includegraphics[scale=0.5]{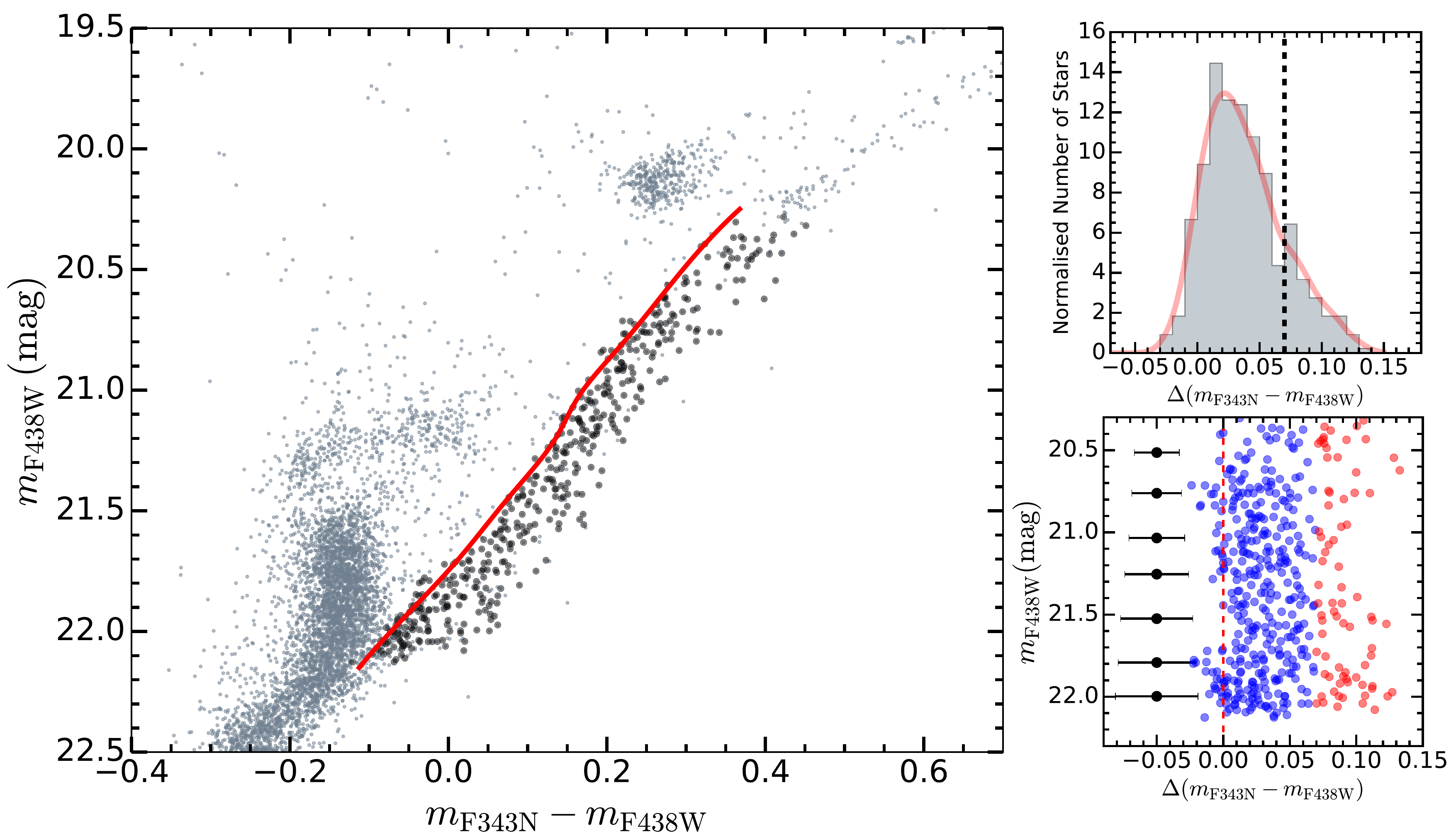}
	\caption{{\it Left Panel}: CMD of NGC 1978 in $Un-B$ vs. $B$ space. Black filled circles indicate the final selected RGB stars. The red solid line marks the fiducial line defined on the blue edge of the RGB. {\it Right Panel}: Histogram of the distribution of RGB stars in NGC 1978 (top), in $\Delta$($Un-B$) colours. The red thick curve represents the KDE. The black vertical dashed line marks the adopted separation for FP and SP stars. $\Delta$($Un-B$) vs. $B$ is shown on the right bottom panel, where blue (red) filled circles denotes FP (SP) stars and the red dashed line marks the verticalised fiducial line. On the left side, the median of photometric errors in colour is reported as black filled circles in bins of 0.25 mag. The errors on $B$ magnitudes are smaller than the marker size.}
	\label{fig:S3}
\end{figure*}

\section{Analysis of NGC 1978}
\label{sec:analysis}

\subsection{The CMD in UV colours}
\label{subsec:cmduv}

Fig. \ref{fig:S2} shows the CMD of NGC 1978 in $U-B$ vs. $B$ (left panel) and $Un-B$ vs. $B$ space (right panel). 
From a first inspection, the CMDs show a visible splitting in the lower RGB, plus a broadened width of the RGB when compared to the observational errors. 
The median of errors in colour and magnitude is reported on the right side of both figures as red filled circles in bins of $\sim$ 0.2 mag. 
We will discuss in the next Section (\S \ref{subsec:rgb}) how we carried out the analysis in order to explain these features as chemical variations of C and N, i.e. as the presence of a second stellar population in the young cluster NGC 1978. We stress here that the observed red sequence in the RGB survived after decontamination from field stars. 

\begin{figure*}
	\centering
	\includegraphics[scale=0.37]{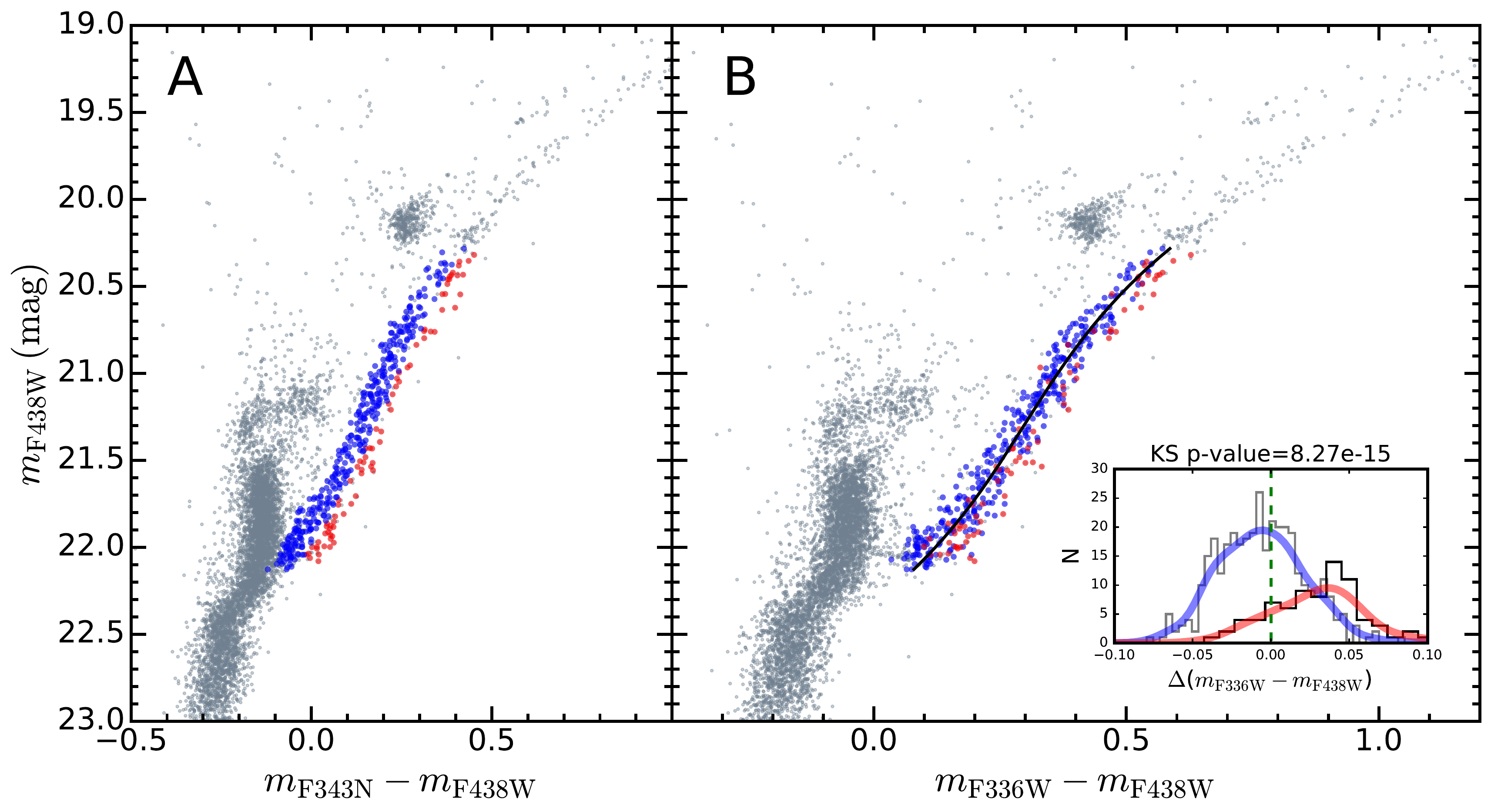}
	\includegraphics[scale=0.37]{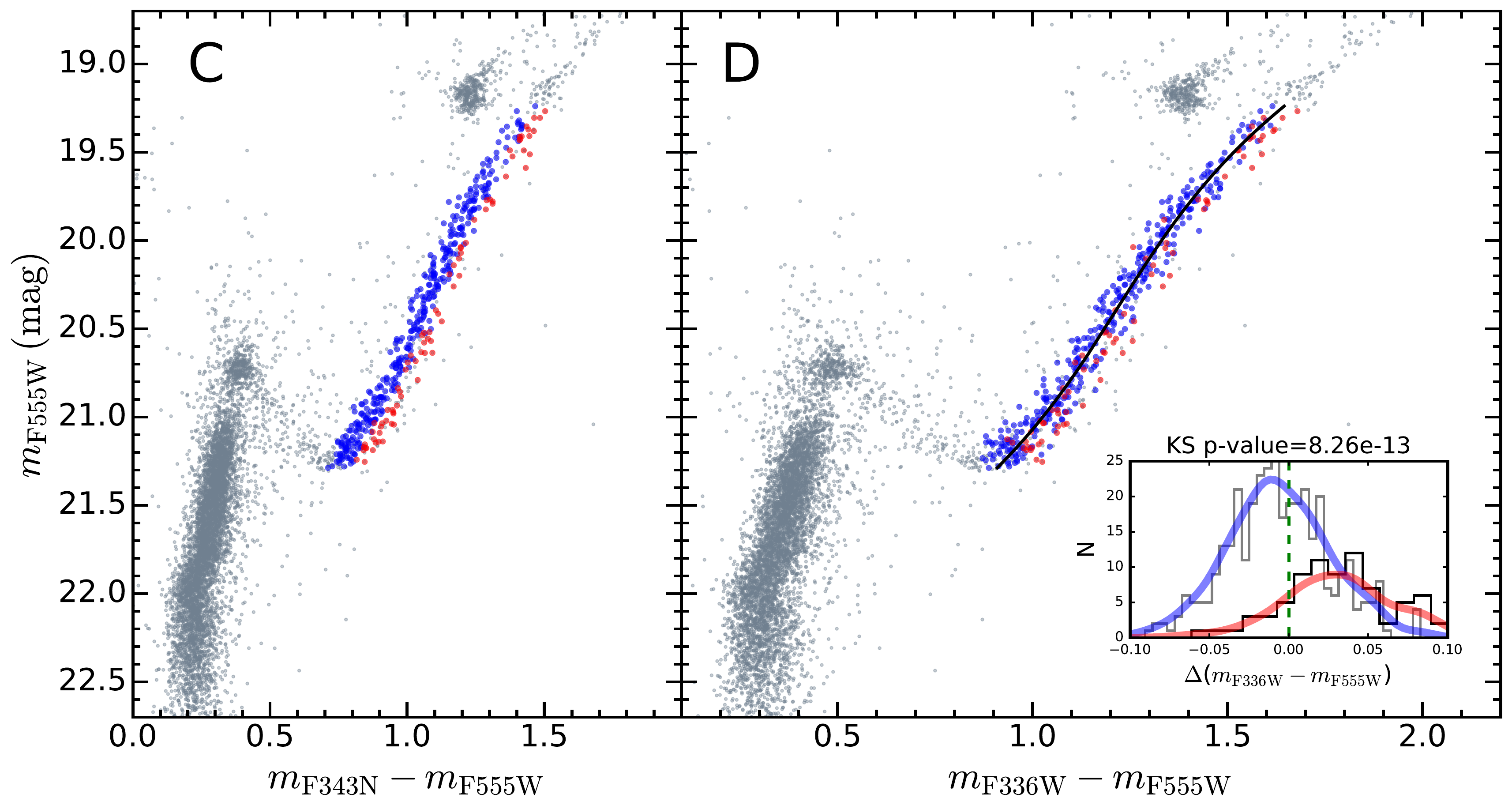}
	\includegraphics[trim={0 0 0.8cm 0}, clip=true, scale=0.37]{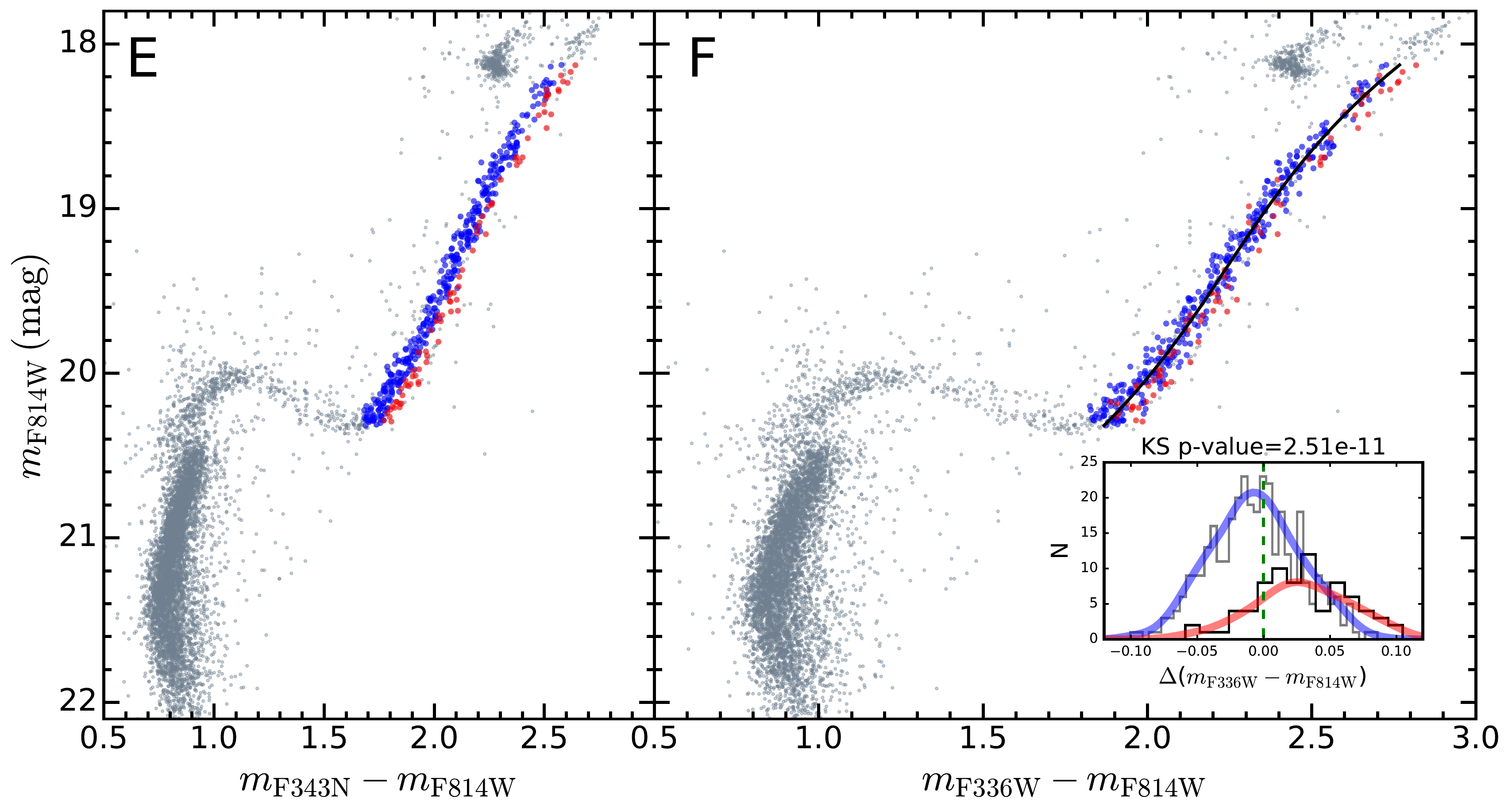}
	\caption{CMDs of NGC 1978 in $Un-B$ vs. $B$ (panel A), $U-B$ vs. $B$ (panel B), $Un-V$ vs. $V$ (panel C), $U-V$ vs. $V$ (panel D), $Un-I$ vs. $I$ (panel E), $U-I$ vs. $I$ (panel F). Blue (red) filled circles mark the selected FP (SP). The black solid lines mark the defined fiducial lines. In the insets of the right panels, the histograms of the distribution of the RGB FP (gray) and SP (black) stars is shown, in $\Delta$($U-B$) (panel B), $\Delta$($U-V$) (panel D), and $\Delta$($U-I$) (panel F) colours. The blue (red) thick curve represents the KDE for the FP (SP) stars. The dashed green line represents the verticalised fiducial line.}
	\label{fig:S4}
\end{figure*}

\subsection{The Red Giant Branch}
\label{subsec:rgb}

Our goal is to conservatively select RGB cluster members, in order to search for evidence of MPs. 
While optical colours are not sensitive to star-to-star C, N, O abundance variations, they can be extremely useful for selecting a clean sample of RGB stars. We thus made the first selection in $B-I$ vs. $I$ CMD and a second one in $V- I$ vs. $I$ CMD.
Finally, we made a last selection in the $Un-B$ vs. $B$ CMD, as a few objects were scattered far off the RGB. Fig. \ref{fig:S3} shows the $Un-B$ vs. $B$ CMD of NGC 1978 with the final selected RGB stars marked as black filled circles.

We defined a fiducial line on the blue edge of the RGB in the $Un-B$ vs $B$ CMD and this is displayed in the left panel of Fig. \ref{fig:S3} as a solid red line.
Next, we calculated the distance in $Un-B$ colours of each RGB star from the fiducial line, \Dunb. We show the histogram of the distribution in \Dunb colours in the top right panel of Fig. \ref{fig:S3}. 
We also derived the kernel density distribution (KDE) from a Gaussian kernel. This is shown in the top right panel of Fig. \ref{fig:S3} as a thick red curve. 
The KDE reveals both broadening and asymmetry in the distribution, along with the presence of a bump for \Dunb $\gtrsim$ 0.06. Hence, we adopted \Dunb $= 0.07$ as the verticalised colour to separate the first population of stars (FP) from the second population (SP). 
This separation is represented with a black vertical dashed line in the top right panel of Fig. \ref{fig:S3}. 
The bottom right panel of Fig. \ref{fig:S3} shows the \Dunb colours vs. $B$, where FP are represented with blue filled circles while SP are represented in red. 
The red dashed vertical line marks the adopted fiducial line. We find that the SP represent 18\% of the total selected RGB stars.

In order to verify that this broadening is due to the presence of multiple populations and not to photometric errors or field stars, we performed several tests. 

\begin{figure}
	\centering
	\includegraphics[trim={0.5cm 0 0.4cm 0}, clip=true,scale=0.39]{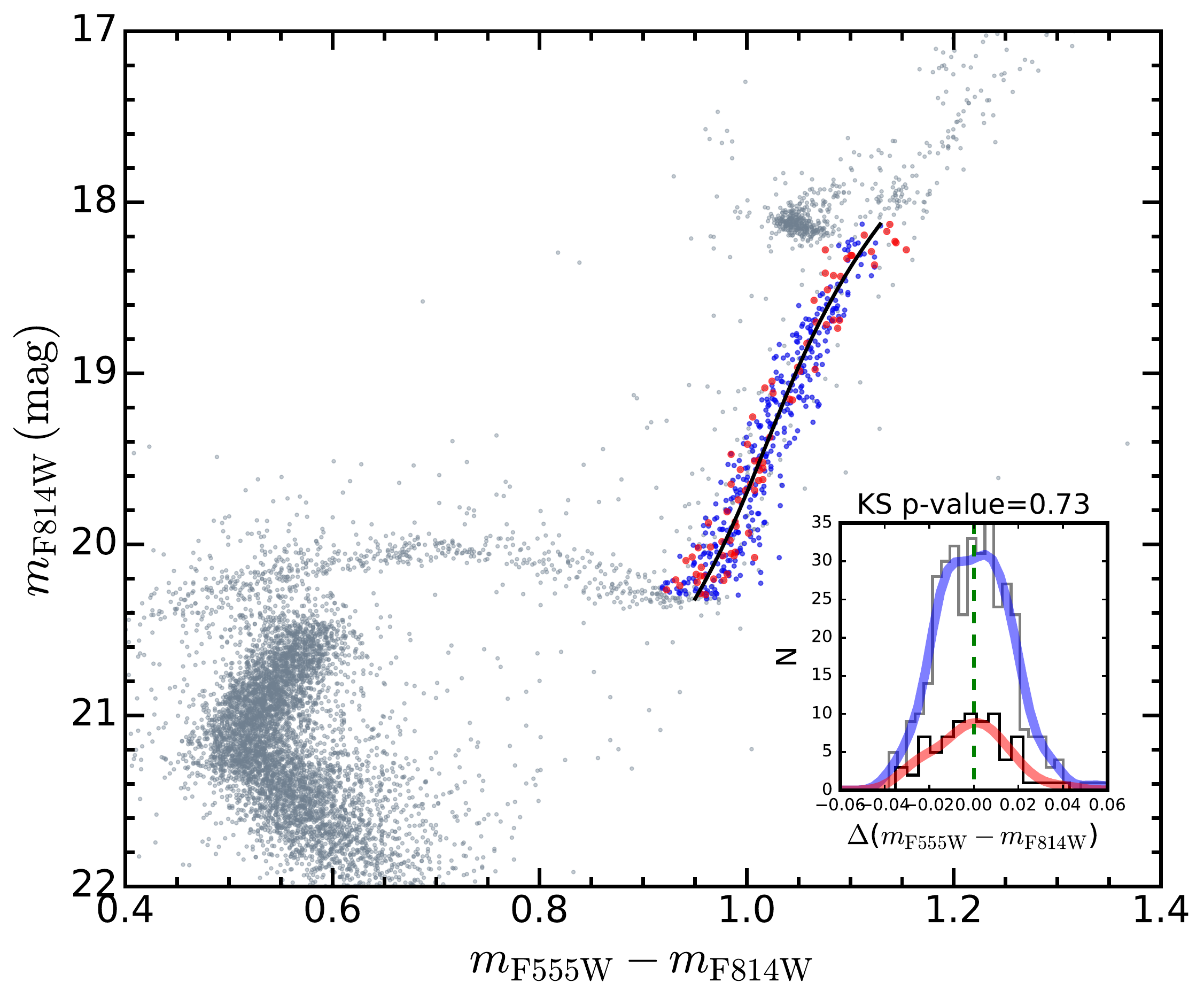}
	\caption{CMD of NGC 1978 in $V-I$ vs. $I$ space. Symbols as in Fig. \ref{fig:S4}. The black solid line marks the defined fiducial line. In the insets of the right panels, the histogram of the distribution of the RGB FP (gray) and SP (black) stars is shown, in $\Delta$($V-I$) colours. The blue (red) thick curve represents the KDE for the FP (SP) stars. The dashed green line represents the verticalised fiducial line.}
	\label{fig:S5}
\end{figure}

As stressed in \S \ref{sec:obs}, we derived our final photometric catalogue by selecting stars based on photometric quality indicators. We additionally checked the quality of our photometry by comparing the average errors of the two populations in $Un-B$ colours, with both having a mean error of $\sim 0.025$. 
Then, we investigated whether a more severe selection based on photometric errors (see \S \ref{sec:obs}) might invalidate our results. 
We selected {\it bona-fide} stars by applying a sigma-rejection in the error versus magnitude diagrams ($U$, $Un$ and $B$) of our original catalogue. 
We derived the median values in bins of 0.5 mag for each diagram and excluded stars in each bin at more than 3$\sigma$ from the median. 
We repeated the same analysis with 1$\sigma$ and 2$\sigma$ cuts. Our results stay unchanged and the SP sequence never disappeared.

In Fig. \ref{fig:S4} we plotted the FP (blue filled circles) and SP (red filled circles) in several colour-magnitude spaces. 
Panel A shows the $Un-B$ vs. $B$ CMD of NGC 1978. SP stars follow a red sequence on the RGB while FP stars follow a blue sequence, as expected from their selection.  
Panel B shows the $U-B$ vs. $B$ CMD, panels C and D show the $Un-V$ vs. $V$ and $U-V$ vs. $V$ and, lastly, panels E and F show the $Un-I$ vs. $I$ and $U-I$ vs. $I$ CMDs, respectively.
We defined a fiducial line for the selected RGB stars in 
the $U-B$ vs $B$, $U-V$ vs $V$ and $U-I$ vs $I$ CMDs. These are displayed as black solid lines in panels B, D and F of Fig. \ref{fig:S4}. We then calculated the distance of each RGB star from the fiducial line to obtain the $\Delta$($U-B$), $\Delta$($U-V$) and $\Delta$($U-I$) verticalised colours.
In the insets of panels B, D and F of Fig. \ref{fig:S4}, the histograms of the distribution in $\Delta$(Colours) of the FP (gray) and SP (black) stars is shown. The blue (red) thick curve represents the KDE for the FP (SP) stars. Note that the FP and SP sequences were selected in the $Un-B$ vs. $B$
CMD.
In all panels, the distribution of FP presents a clear offset from the SP, with the FP sequence bluer and the SP redder. 
In the right panels, this was specifically highlighted by showing the $\Delta$(Colors) distributions, where the peak of the FP KDE is clearly shifted with respect to the peak of the SP KDE.
To quantify this, we performed the Kolmogorov-Smirnov (KS) test between the FP and SP stars distributions in the three $\Delta$(Colours), in order to understand if they are consistent with having been sampled from the same parent distribution. We derived {\it p-values} $<10^{-10}$ for all three $\Delta$(Colours), thus demonstrating that the distributions are indeed different.
This clearly points toward a chemical effect, as errors on $B$ (WFC3/UVIS) are completely independent from errors on $V$ and $I$ (ACS/WFC).  We note that the selection of FP and SP stars was made in the $Un-B$ vs. $B$ CMD so the $U-V$ vs. $V$ and $U-I$ vs. $I$ CMD are completely independent measurements.

\begin{figure}
	\centering
	\includegraphics[scale=0.45]{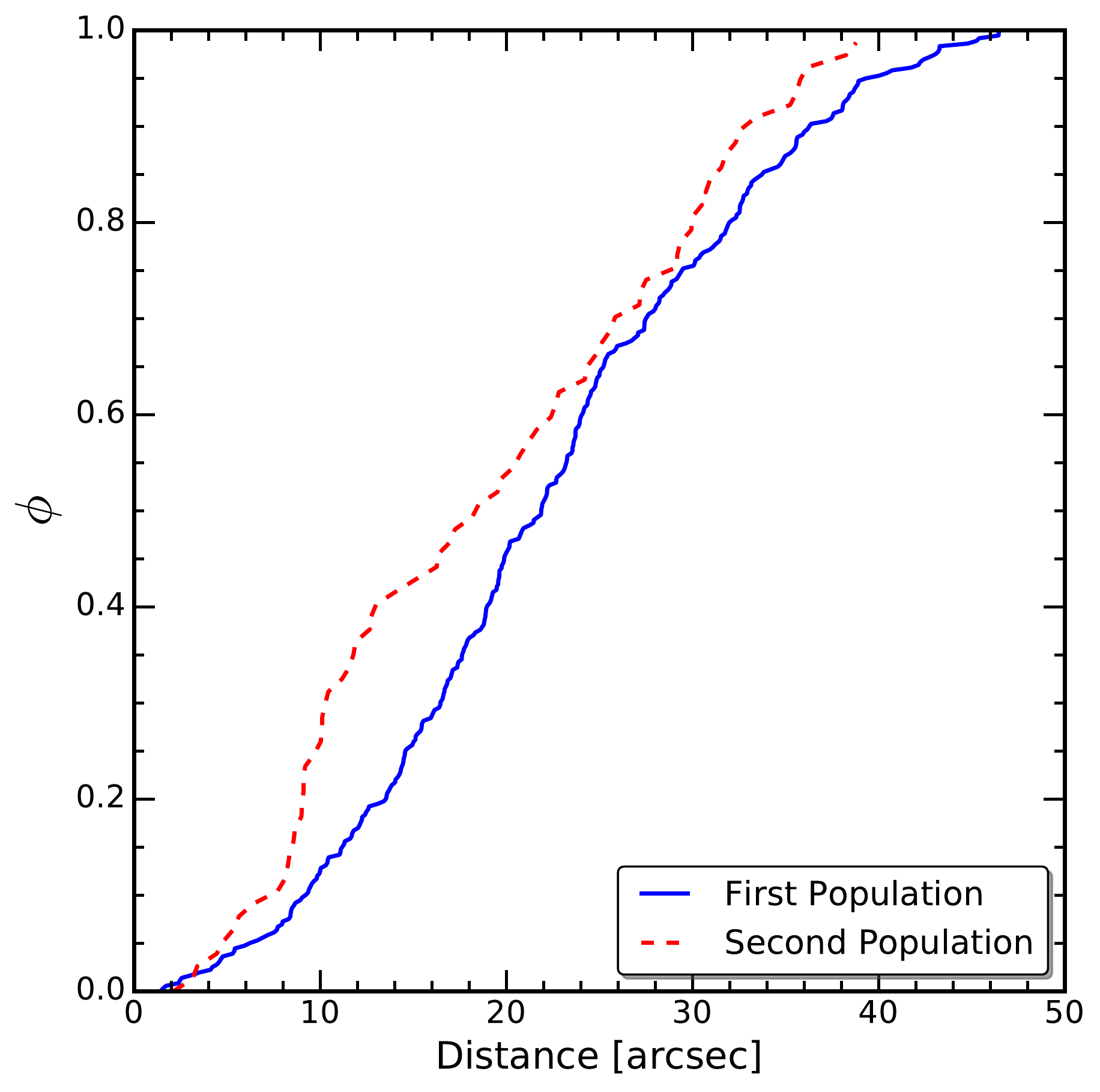}
	\caption{Cumulative distribution of FP (blue solid line) and SP (red dashed line) as a function of the radial distance from the cluster centre.}
	\label{fig:S6}
\end{figure}

\begin{figure*}
	\centering
	\includegraphics[scale=1.]{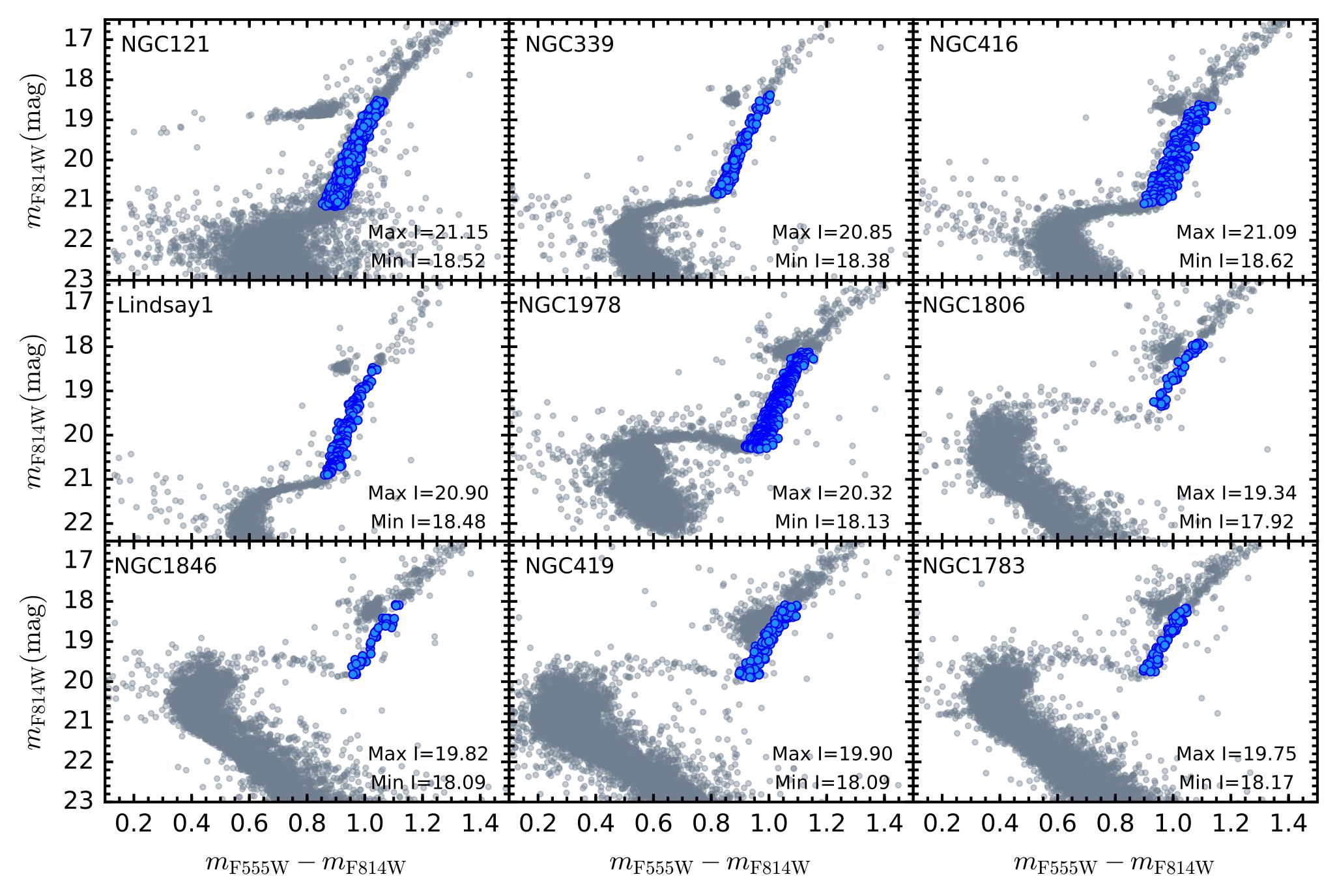}
	\caption{$V-I$ vs. $I$ CMDs for all the targeted clusters in our HST survey. Blue filled circles indicate the selected RGB stars for each cluster. The maximum and minimum values of the selection in $I$ magnitudes are superimposed in each panel.}
	\label{fig:S7}
\end{figure*}

Fig. \ref{fig:S5} shows the $V-I$ vs. $I$ CMD of NGC 1978 with FP (SP) superimposed as blue (red) filled circles. 
We performed the same analysis as in the UV CMDs (see Fig. \ref{fig:S4}) and we show the histograms of the distribution for FP (in gray, blue KDE) and SP (black, red KDE) stars in $\Delta$($V-I$) colours in the inset of Fig. \ref{fig:S5}.
In this combination, as expected given the lack of sensitivity to MP in these filters, , the two sequences are well mixed, showing no signs of systematic offsets. This is confirmed by the KS test, which returns a {\it p-value} $=0.73$.

Finally, we report on how the two populations are distributed as a function of radial distance from the centre of the cluster. Fig. \ref{fig:S6} shows the normalized cumulative radial distribution of the primordial (blue solid line) and enriched (red dashed line) populations. We found that the second population is more centrally concentrated than the FP up to a distance of $\sim$ 50'' from the cluster centre. We performed the KS test on the two population radial distributions. The probability that they belong to the same parent distribution is relatively low, being $<$ 1\%. 
This test statistically shows that the two radial distributions are likely different. Additionally, the evidence for the SP to be centrally concentrated reinforce the cluster membership of the red sequence stars. 
We also corrected our photometric dataset for differential reddening (DR) and we repeated the exact same analysis. All results stay unchanged. For more specific details on the DR correction procedure, we refer interested readers to \cite{paperII, paperIII}. 

It might be argued that the SP sequence is composed of evolved binaries on the RGB, due to the likely presence of a large number of blue stragglers in NGC 1978 (see Fig. \ref{fig:S1}).
However, evolved blue straggler stars (BSS) should instead be slightly bluer in colour than the RGB, according to \cite{sills09}, where they assumed that blue stragglers were formed through stellar collision. This is not compatible to what we observe, as the SP distribution of stars is well overlapped to the FP sequence in the $V-I$ vs $I$ CMD (Fig. \ref{fig:S5}).
Additionally, \cite{tian06} looked at BSS formed via mass transfer rather than collision. As pointed out by \cite{sills09}, their evolutionary tracks of the BSS are comparable to normal stars after the mass transfer occurred. Thus, the same condition mentioned above must apply.

All tests point towards the conclusion that MPs are detected in the RGB of NGC 1978. To date, this is the youngest cluster hosting chemical abundance anomalies. 

The same analysis reported in this Section was performed on the younger clusters in our sample (namely, NGC 419, 1783, 1806 and 1846). 
On the contrary, we found no clear evidence for MPs in their RGBs. The detailed analysis 
of the younger clusters will be reported in a forthcoming paper. We will report the global results from our HST survey in the next Section (\S \ref{sec:results}).

\section{HST Survey Results}
\label{sec:results}

The CUBI pseudo-colour (defined as ($U-B$)$-$($B-I$)) has been shown to be very effective at unveiling multiple sequences on the RGB \citep{dalessandro16, monelli13}. Here, we take advantage of a similar pseudo-colour, known as CUnBI $\equiv$($Un-B$)$-$($B-I$), to provide a direct and quantifiable comparison between all the clusters in our survey. Additionally, these pseudo-colours do not have a strong dependence on the effective temperature of stars, such that the RGB is almost vertical in the diagram, i.e. splittings are more easily discernible in this combination and are also largely independent of potential spreads in He \citep{paperIII}. We finally exploit CUnBI to compare our data with stellar atmosphere models in \S \ref{sec:models}. 

\begin{figure*}
	\centering
	\includegraphics[scale=0.45]{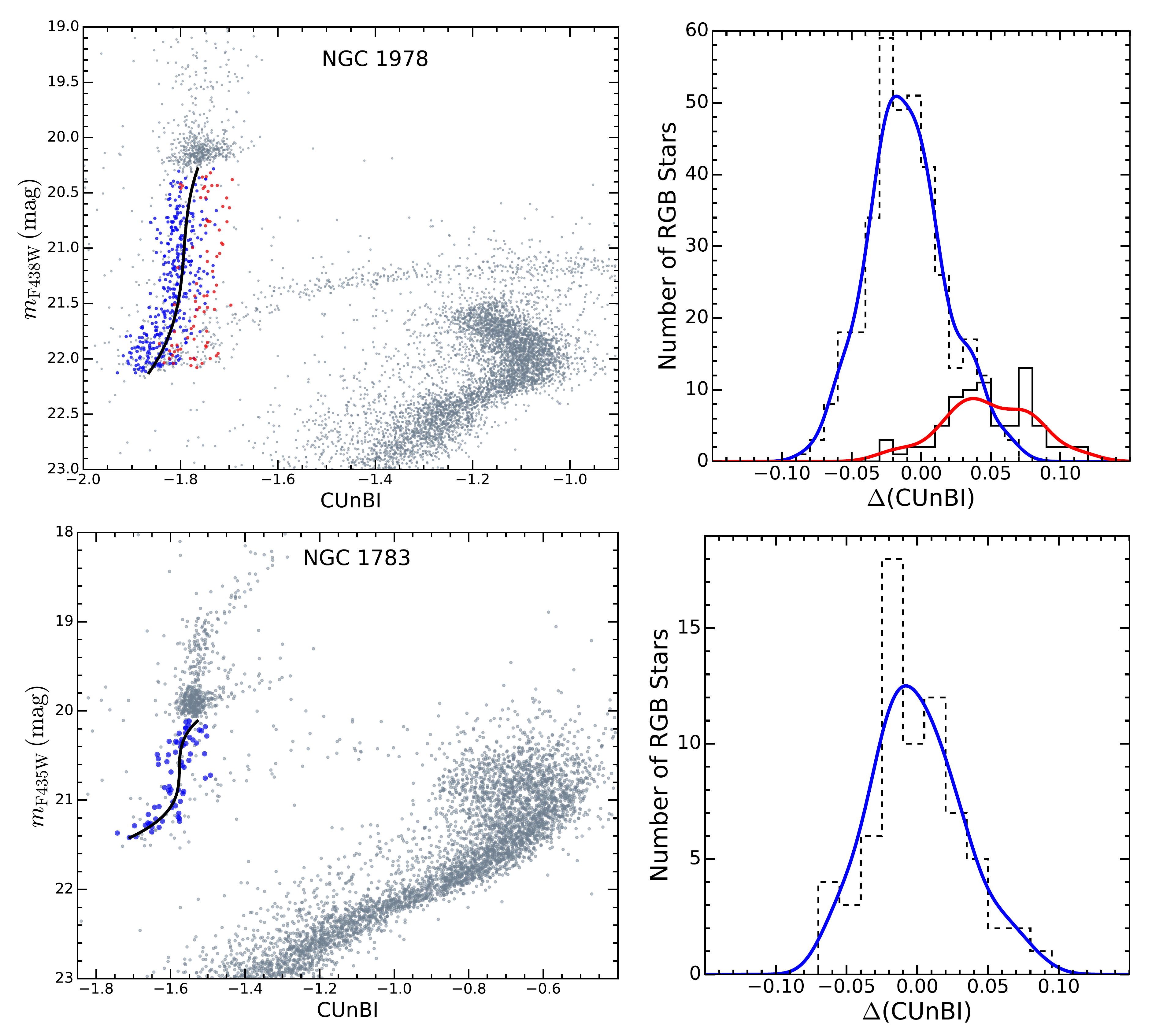}
	\caption{Top (bottom) left panels: CunBI vs $B$($m_{\rm F435W}$) CMD of NGC 1978 (NGC 1783), where symbols are as in Fig. \ref{fig:S4}. The black solid line indicates the fiducial line. Top (bottom) right panels: histogram of the distributions in $\Delta$(CUnBI) colours for the FP (dashed) and SP (solid) for NGC 1978 (NGC 1783), with KDEs superimposed (blue for the FP and red for the SP). }
	\label{fig:S8}
\end{figure*} 

We adopted the same analysis for each cluster. RGB stars have been selected in three CMDs of each cluster to establish membership. 
Figure \ref{fig:S7} shows the $V-I$ vs. $I$ CMDs for all the targeted clusters in our survey. Blue filled circles represent the RGB stars which passed all three selections. 
We defined the selection by choosing stars in the lower RGB, which are fainter than the RGB bump (to avoid chemical mixing due to stellar evolutionary effects). 
We verticalised our selected RGB stars in CUnBI vs $B$ CMD and we calculated the distance  $\Delta$CUnBI from the fiducial line for each star. The top left panel of Fig. \ref{fig:S8} shows the CUnBI vs. $B$ CMD of NGC 1978, while the bottom left panel shows the CUnBI vs. $B$ CMD of NGC 1783, where MPs were not detected. Symbols are as in Fig. \ref{fig:S4}. The black solid line marks the defined fiducial line to derive the $\Delta$CUnBI. As expected in this filter combination, for NGC 1978, FP and SP stars show an unambiguous offset in colours, with the FP having bluer colours and the SP redder. 
This is also evident in the right top panel of Fig. \ref{fig:S8}, where the NGC 1978 histogram of the $\Delta$CUnBI distributions for both FP and SP is shown. The KDE distributions are also indicated, in blue for the FP and in red for the SP. The same is shown on the bottom right panel of Fig. \ref{fig:S8}, but for NGC 1783, where there is no hint for asymmetry or broadening. 

\begin{figure*}
	\centering
	\includegraphics[scale=0.45]{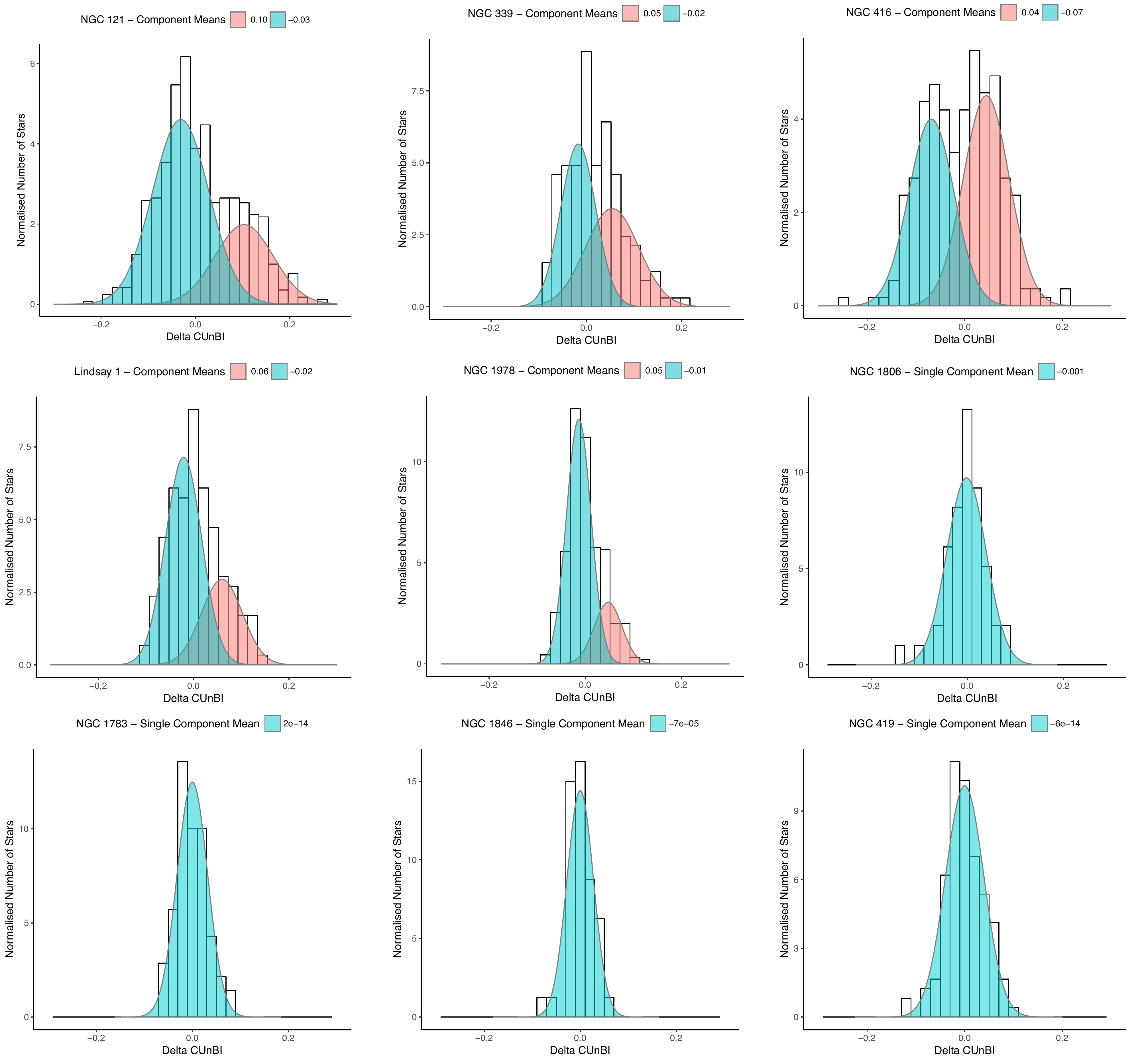}
	\caption{Histograms of the distribution of RGB stars in CUnBI colours for all clusters in our survey. The GMM fit on unbinned data are marked in blue for the FP and red for the SP. The means of the Gaussian distributions are labelled on top of the plots. }
	\label{fig:S9}
\end{figure*}

Next, we fit the unbinned verticalised $\Delta$CUnBI data with Gaussian Mixture Models (GMMs, see Fig. \ref{fig:S9}, where we also plot the binned $\Delta$CUnBI distribution for visual representation) to identify the presence of multiple Gaussian components in the colour distribution, hence two or more populations with different N abundance. 
We fit the data with the SCIKIT-LEARN python package called MIXTURE\footnote{\url{http://scikit-learn.org/stable/modules/mixture.html}}, which applies the expectation-maximization algorithm for fitting mixture-of-Gaussian models. 
In order to determine the number of Gaussians which best reproduce the data, we adopted the Akaike Information Criterion (AIC, \citealt{akaike74}). 
We found that the $\Delta$CUnBI distributions of NGC 1978, NGC 416, NGC 339, NGC 121 and Lindsay 1 are best represented with 2 components, while only one component is found for NGC 419, 1783, 1806 and 1846 (see Fig. \ref{fig:S9}).

The final result is that all clusters older than 2 Gyr show multiple Gaussian components in the fit, i.e. MPs are present, while all clusters younger do not. We then defined \drgb as the difference between the means of the two main Gaussian components in the $\Delta$CUnBI distribution. 
This gives a robust indication on the level of N enrichment present in clusters which host MPs. Errors on \drgb were calculated with a bootstrap technique based on 5000 GMMs realizations. 
We set \drgb of clusters which do not show MPs to zero.

Although statistically less likely, to be as conservative as possible, we forced the GMM procedure to fit the one-gaussian component data with two Gaussian distributions. For NGC 1806 and NGC 1846 the fit yields two almost overlapped gaussians, resulting in \drgb $<$ 0.01. In these two cases, we considered the standard deviation of the single Gaussian as upper error on \drgb. For NGC 419 and NGC 1783 the fit finds a separated second Gaussian component, with \drgb $\sim$ 0.04, which was used as upper error on \drgb . In order to establish the statistical significance on the second Gaussian component for NGC 419 and NGC 1783, we calculated the errors on the normalisation factor with 5000 bootstrap realisations. The normalisation was found to be consistent with zero for both clusters. 
Finally, in Table \ref{tab:signi}, we also report for each cluster the probability that a bimodal distribution is rejected, which is in agreement with previous studies \citep{paperI,paperII,paperIII}. This was obtained with a parametric bootstrap technique by using the GMM code by \cite{gnedin10}.

It is worth noting that the GMM fitting method was {\it not} adopted in order to demonstrate that MPs are detected or not in a given cluster. It was just used as a confirmation of our findings. MPs are primarily detected by selecting samples in the $Un-B$ vs. $B$ CMD and by examining where they order in the different filter combinations.
\begin{figure}
	\centering
	\includegraphics[scale=0.36]{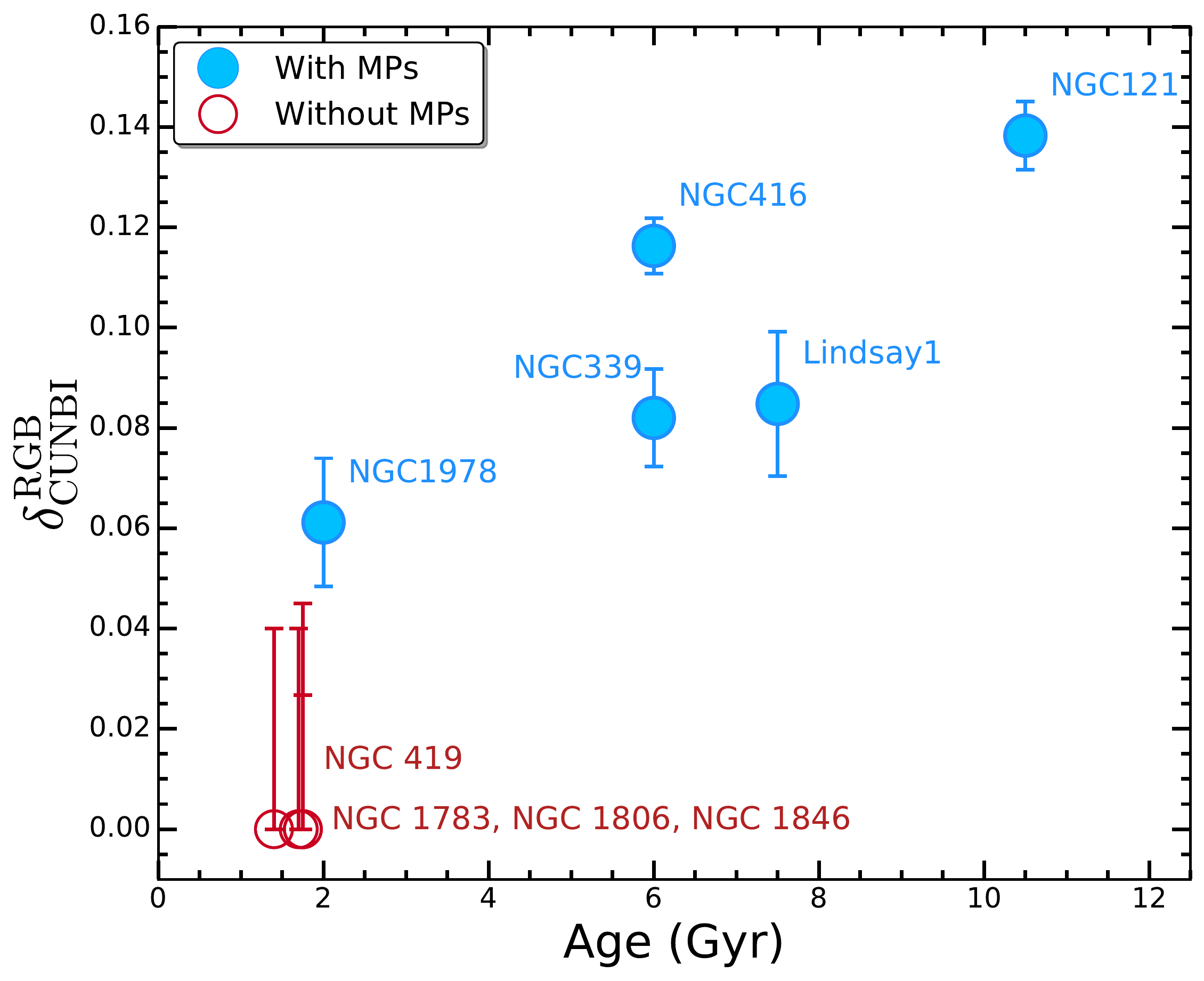}
	\caption{$\delta^{\rm RGB}$(CUnBI) as a function of age for the 9 clusters in our HST photometric survey. Clusters which are found to host MPs are indicated with blue filled circles while clusters which do not show MPs are marked with red open circles.}
	\label{fig:deltacunbiage}
\end{figure}

We plot \drgb as a function of cluster age in Fig. \ref{fig:deltacunbiage} for all our targets. Deviations from zero, indicating N enrichment, are observed for all clusters older than $\sim$ 6 Gyr, confirming our earlier results \citep{paperI,paperII,hollyhead17}. Conversely, the narrow RGB in the CUnBI vs. $B$ CMD would exclude the presence of significant nitrogen star-to-star variations in the youngest clusters of our sample, namely NGC 419, 1783, 1806, 1846, with ages of $\sim 1.5-1.7$ Gyr (see also \citealt{paperIII}). 
Finally, strong evidence for MPs is also found for the first time in the 2 Gyr old cluster, NGC 1978. This finding suggests that age is playing a pivotal role in controlling the presence of MPs, as chemical anomalies are detected in all massive clusters older than 2 Gyr while they are lacking at younger ages.  

However, it is possible that a certain level of N enrichment (smaller compared to the older clusters) is still present in the younger clusters, which cannot be seen due to observational uncertainties and contamination from field stars (see  \S \ref{subsec:comp} and Martocchia et al. in prep.).  Despite this, the main results from this work stay unchanged: we report an unexpected age effect on the onset of MPs. We note that if N-spreads like those observed in the ancient GCs (or those in the 2-8 Gyr old clusters) were present in the clusters younger than 2 Gyr, they would have been readily detected.

\begin{table}
\caption{Columns give the following information: (1) cluster name, (2) cluster age in Gyr, (3) reference for cluster ages, (4) probability that a bimodal $\Delta$CUnBI distribution is rejected for each cluster in the survey.}
\label{tab:signi}
\centering          
    \begin{tabular}{c c c c}     
        \toprule\toprule
        Cluster Name & Age (Gyr) & Ref. & $\Delta$CUnBI pvalue\\
        (1) & (2) & (3) & (4)\\
        \midrule
        NGC 121 & 10.5  & (1) & $<$0.001 \\
        Lindsay 1 & 7.50 &  (2) & 0.16 \\
       NGC 339   & 6.00 & (2) & 0.04 \\
       NGC 416   & 6.00 & (2) & 0.03 \\
       NGC 1978 & 1.90 & (3) & $<$0.001\\
       NGC 1783 & 1.75 & (4) & 0.58\\
       NGC 1846 & 1.75 & (4) & 0.63\\
       NGC 1806 & 1.70 & (4) & 0.51 \\
       NGC 419 & 1.50  & (2) & 0.47 \\
       \bottomrule\bottomrule
        \end{tabular}
        \\
        (1)~\citet{mclaughlin05}; 
        (2)~\citet{glatt08}; 
        (3)~\citet{mucciarelli07};
        (4)~\citet{niederhofer16}.
\end{table}

\section{Models for the Chemical Anomalies}
\label{sec:models}

In this Section, we describe how we calculated stellar atmosphere models with different levels of chemical enrichment and we report comparisons with the observations.

\subsection{Cluster age, metallicity and extinction}
\label{subsec:agemet}

We compared MIST isochrones \citep{dotter16, choi16} to our data to obtain an estimate of age, metallicity, distance modulus ($M-m$) and extinction values. 
For NGC 1978, we found that the MIST parameters which best match the data and the isochrones are: (i) $\log$(t/yr) $= 9.35$ (t $\sim$ 2.2 Gyr), (ii) [Fe/H] $= -0.5$, (iii) $M-m = 18.5$, (iv) $E(B-V)$ = 0.07. The selected isochrone is superimposed on the $V-I$ vs. $V$ CMD of NGC 1978 in Fig. \ref{fig:S1} as a red solid curve. \citep{mucciarelli07} report an age of t = $1.9\pm0.1$ Gyr for NGC 1978 by using several sets of different isochrones (Padua, BaSTI, PEL). Our slightly different choice of age might be due to the different set of isochrones used in this work, i.e. MIST isochrones. 
However, with this work we do not aim at providing an improved estimate of the age of NGC 1978. We rather need to obtain a reliable MIST isochrone fit to our CMD data, as we will use these values as input for MIST models to develop our synthetic photometry. However, as it is clear from Fig. \ref{fig:S1}, our isochrone does not perfectly match data on the Main Sequence or the Main-Sequence Turnoff (MSTO). Indeed, in adopting MIST isochrones, we were not able to match perfectly data for all evolutionary stages with any parameters combination. Nonetheless, this issue is somewhat irrelevant for our studies, since we will focus on the RGB stage. 

\subsection{Comparison with stellar models}
\label{subsec:comp}

We calculated synthetic photometry from model atmospheres with different levels of chemical enrichment, in order to compare them with our photometric data. 
We used version 1.0 of the MIST isochrones \citep{choi16} with an age of 2.2 Gyr and a metallicity of [Fe/H] $= -0.5$ to provide input parameters for our model atmospheres (see \S \ref{subsec:agemet}). Model atmospheres and synthetized spectra were calculated with ATLAS12 \citep{kurucz70, kurucz05} and SYNTHE \citep{kurucz79, kurucz81}. We used (i) line lists for the atomic data as \cite{larsen12} and \cite{larsen14}, (ii) PYTHON wrappers for ATLAS12 and SYNTHE as used by \cite{larsen12}, and (iii) the solar abundances adopted by \cite{asplund09} for the stellar atmospheres calculations. 
We refer the interested reader to \cite{paperIII}, for more details about the models.

 We adopted three chemical mixtures. First, we calculated a set of scaled solar models ([C/Fe] $=$ [N/Fe] $=$ [O/Fe] $=$ 0). Next, we calculated a set of N-enhanced models with [C/Fe] $=$ [O/Fe] $= -0.1$ dex and [N/Fe] $= +0.5$ dex. Lastly, we calculated a set of models with slightly less degree of N-enhancement but with solar C and O abundances ([C/Fe] $=$ [O/Fe] $=$ 0 and [N/Fe] $= +0.3$ dex), in order to check degeneracies between C depletion and N enhancement in CUnBI colours.
 
The C and O abundances were chosen to keep the [(C+N+O)/Fe] constant between the models, according to what we observe in standard GCs \citep{brown91, cohen05, yong15, marino16}.
For each of these chemical mixtures we kept the helium abundance (surface Y = 0.248) constant and all other abundances fixed at solar. We also assumed that the model atmospheres had the same chemical abundances at all stellar evolutionary stages.
We then integrated our model spectra over the filter transmission curves for WFC3\footnote{\url{http://www.stsci.edu/hst/wfc3/ins\_performance/throughputs/Throughput\_Tables}} and ACS/WFC\footnote{\url{http://www.stsci.edu/hst/acs/analysis/throughputs}} and used the zeropoints provided on each instrument’s website to calculate Vega magnitudes. We directly compare our models to the data. 

\begin{figure}
	\centering
	\includegraphics[scale=0.31]{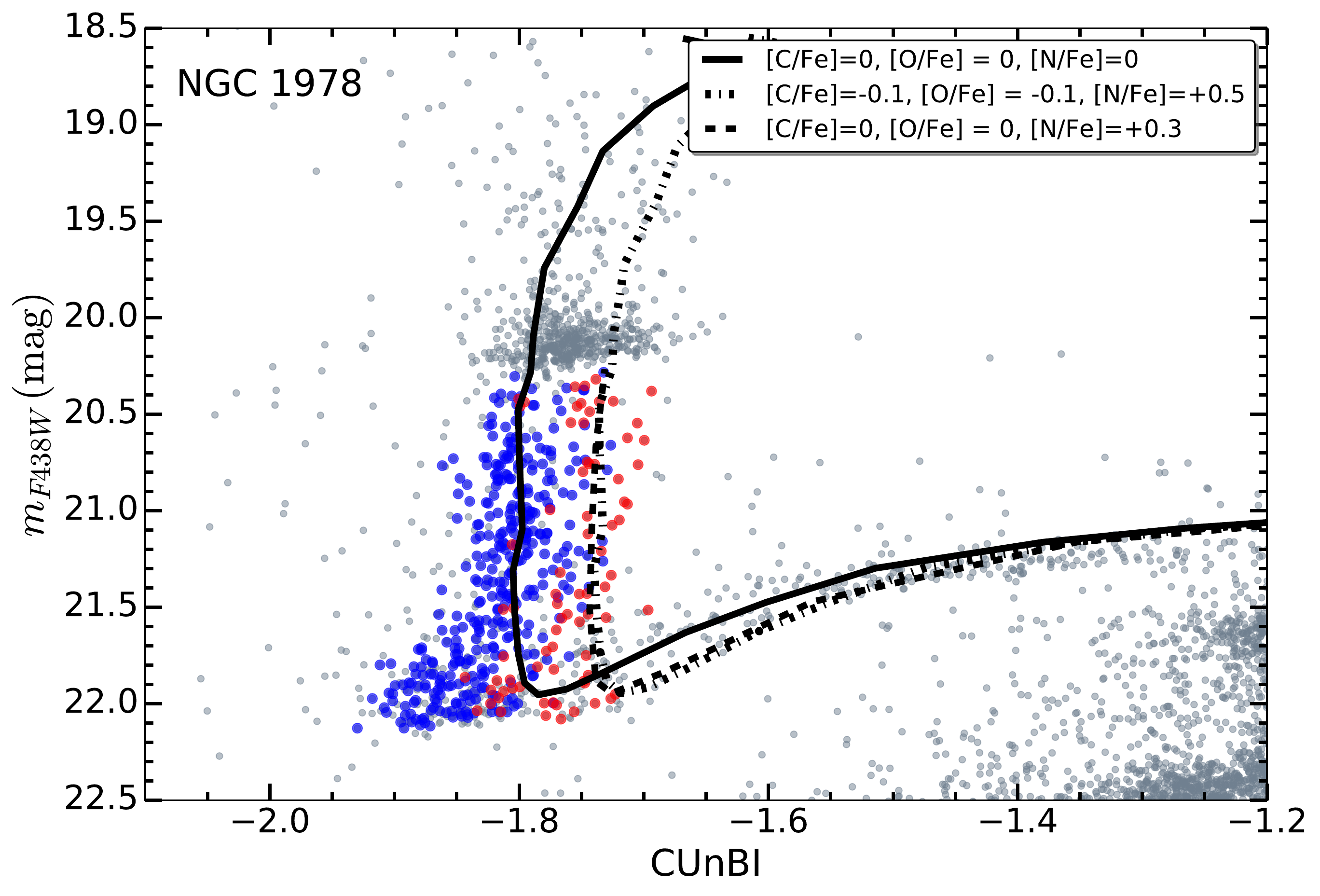}
	\caption{CUnBI vs. $B$ CMD of NGC 1978 zoomed in the RGB region. Blue (red) filled circles indicate the FP (SP) stars. Black solid, dash-dotted, dashed curves represent stellar isochrones for [C/Fe] $=$ [N/Fe] $=$ [O/Fe] $= 0$, [C/Fe] $=$ [O/Fe] $= -0.1$ and [N/Fe] $= +0.5$, and [C/Fe] $=$ [O/Fe] $= 0$ and [N/Fe] $= +0.3$ chemical abundance mixtures, respectively.}
	\label{fig:S10}
\end{figure}

We exploited the CUnBI pseudo-colour to give an estimate of the level of nitrogen enrichment for the SP stars when MPs are detected and to provide an upper limit where MPs are not detected. Fig. \ref{fig:S10} shows the CUnBI vs. $B$ CMD of NGC 1978, with a zoom in the RGB region. The black solid, dash-dotted, dashed curves represent isochrones for the three chemical mixtures described above, respectively. Blue (red) filled circles mark the selected FP (SP) RGB stars.
Fig. \ref{fig:S10} shows that the [N/Fe]$=+0.5$ dex and [N/Fe]$=+0.3$ dex models are almost completely overlapped, with [C/Fe] and [O/Fe] abundances depleted in the former but kept solar in the latter. This means that we expect the same spread on the RGB with a different combination of C and N, i.e. there is a degeneracy between C and N abundances. Unfortunately, a reliable measure of N enrichment by comparing the observed CUnBI vs. $B$ CMDs with the models cannot be given, as no spectroscopic measurements of chemical abundances are currently available for this cluster. 
However, we can still provide a rough estimate, by assuming that MPs in younger clusters are the same as those of ancient GCs, i.e. an enhancement in N is associated with a depletion in C \citep{cannon98}. 

Accordingly, we report a N enrichment for NGC 1978 of [N/Fe] $\sim +0.5$ dex. Also, we can make a differential comparison among NGC 1978 and the younger clusters. 
We do not detect MPs in NGC 419, 1783, 1806 and 1846 but we cannot exclude the presence of a slight N variation due to observational uncertainties and decontamination from field stars (see \S \ref{sec:results}). If we assume a [N/Fe] $\sim +0.5$ dex for NGC 1978, then we can set an upper limit to any enrichment in the younger clusters (NGC 419, 1783, 1806, 1846) of [N/Fe]$<+0.3$ dex. 

\begin{figure*}
	\centering
	\includegraphics[scale=0.8]{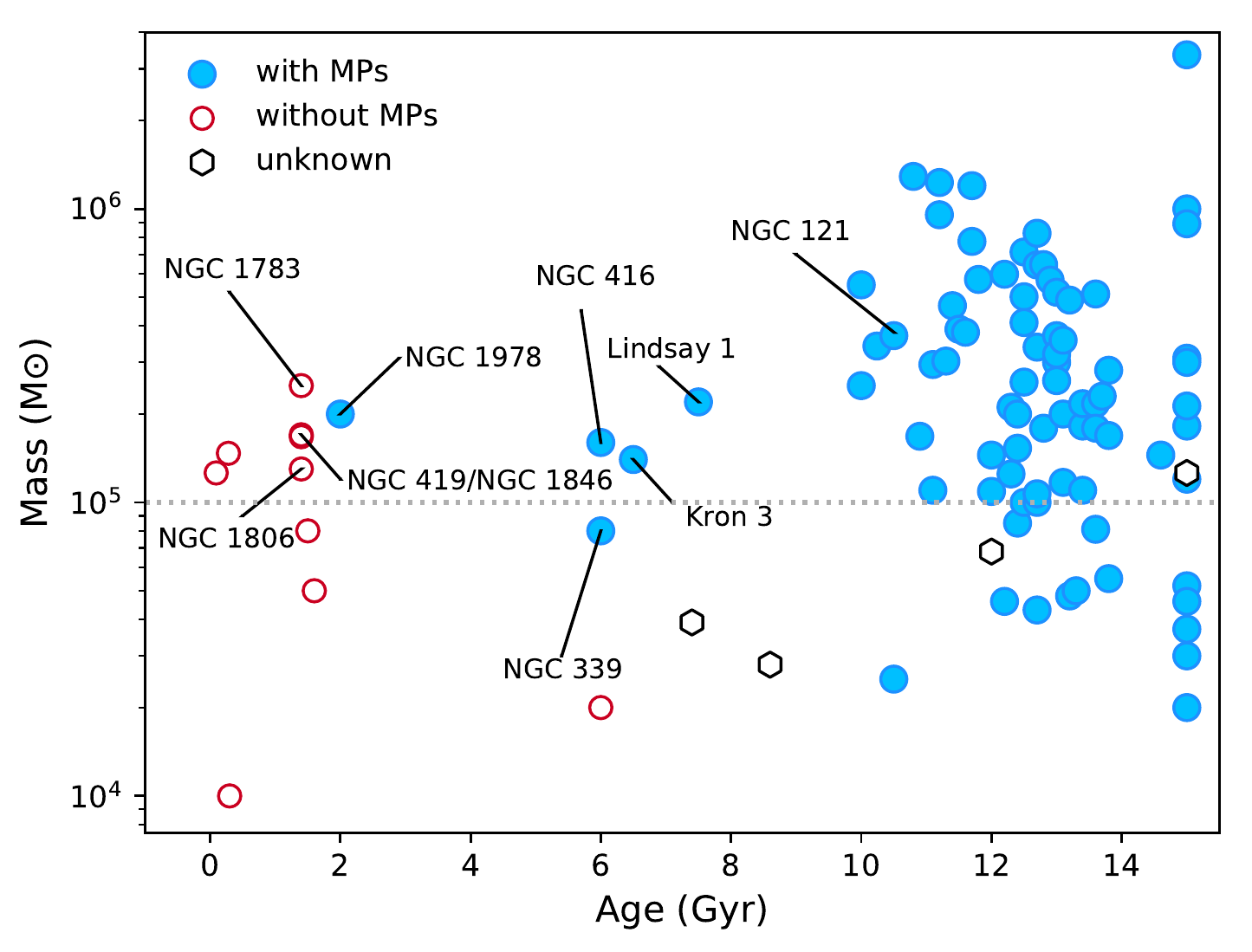}
	\caption{The relation between age and mass for the clusters in our survey (labelled) as well as clusters taken from the literature. Clusters which are found to host MPs are indicated with blue filled circles while clusters which do not show MPs are marked with red open circles.  Clusters where the presence or absence of MPs is still under debate are shown as open hexagons.}
	\label{fig:ageM}
\end{figure*}

\section{Discussion and Conclusions}
\label{sec:discussion}

In this paper, we reported the photometric analysis of new and archival HST images for the LMC cluster NGC 1978 as well as the results from the remaining clusters in our sample.
For NGC 1978, the $U-B$ vs. $B$ and $Un-B$ vs $B$ CMDs revealed hints of a splitting in the lower RGB. Additionally, the RGB showed a broadened width compared to the observational errors (Figs. \ref{fig:S2} and \ref{fig:S3}).
We selected RGB stars in three different CMDs and used the $Un-B$ vs. $B$ CMD to separate the FP from the SP stars. We plotted the FP and SP sequences in several UV CMDs and in all 
of them, the distribution of FP shows an offset from the SP, with the FP sequence bluer and the SP redder, on average (Fig. \ref{fig:S4}). This does not occur when considering optical CMDs (Fig. \ref{fig:S5}). We also showed that the SP stars present in NGC 1978 is more centrally concentrated than the FP stars (Fig. \ref{fig:S6}). This strongly confirms cluster membership for the red-sequence stars, as well as the likelihood that they are indeed the product of separate chemical evolution from the FP stars, as SP stars are almost always found to be more centrally concentrated than FP stars.

We can then conclude that MPs are detected in the RGB of NGC 1978, which is $\sim$ 2 Gyr old and has a mass of $2 -4  \times 10^{5}$ \msun. This is the youngest cluster found to host star-to-star 
abundance variations so far. Future observations to determine if Na (and Al, O, etc...) are necessary to further quantify the results presented here. However, we note that in older GCs where MPs are confirmed to be present, the observed Na spreads are much smaller than the N spreads, hence in NGC~1978 we may only expect Na spreads of $\sim0.2$~dex or less. We also revealed that MPs are not instead detected in the RGB of the younger clusters in our survey, namely NGC 1783, 1806, 1846. We will present their detailed analysis in a forthcoming paper. 

We then presented the global results from our HST photometric survey. We took advantage of the CUnBI colour combination in order to compare all the clusters in our survey(\S \ref{sec:results} and Figs. \ref{fig:S9}, \ref{fig:deltacunbiage}). All clusters older than 2 Gyr host MPs, while chemical anomalies are not detected in those younger than this age. 
This firmly suggests that there is an age effect in the onset of multiple populations.

We also developed stellar atmosphere models for the chemical anomalies. We exploited CUnBI to give a rough estimate of the level of N enrichment for the SP stars when MPs are detected and to provide an upper limit where MPs are not detected. We reported a N enrichment for NGC 1978 of [N/Fe] $\sim +0.5$ dex. Comparing NGC 1978 with the younger clusters, we can set an upper limit to any enrichment in the younger clusters (NGC 419, 1783, 1806, 1846) of [N/Fe]$<+0.3$ dex (\S \ref{sec:models}). 

In Fig. \ref{fig:ageM} we plot a summary of our HST survey as well as results taken from the literature. 
We also add here the results for the $\sim$6.5 Gyr old SMC cluster Kron 3 (Hollyhead et al. 2017, submitted), where MPs have been spectroscopically identified. 
All of the clusters in our HST survey have masses in excess of $\sim10^5$ \msun, e.g. the mass for which MPs are almost always found in ancient GCs, but we detect MPs only in clusters older than $\sim$2 Gyr. 

Taken at face value, this suggests that some mechanism operating only in stars less massive than 1.5 \msun (the mass of a RGB star at $\sim$2 Gyr) may be responsible for the onset of MPs. 
Note in Fig. \ref{fig:ageM} that there is one cluster older than 2 Gyr where MPs were not detected. This is Berkeley 39, which is a low-mass cluster ($\sim10^4$ \msun). Clearly, age is not the only controlling parameter, cluster mass must also come into play, as well as environment might contribute. Future work will be needed to consider the impact of environment on MPs, as a significant factor shaping the amount of chemical anomalies in clusters. Conversely, Ruprecht 106 and IC 4499 are two old clusters ($>12$ Gyr) where the presence of MPs is still under debate, with preliminary studies suggesting that MPs may not be present \citep{villanova13,walker11}. Also, the presence of MPs is still unknown for other two relatively old clusters, namely Terzan 7 and Pal 12 ($7-9$ Gyr), as only a few stars (less than 5) have light element abundance measurements (e.g. \citealt{cohen04}).

The dependence of light element variations on age is not predicted by any model that has been proposed to explain the formation and evolution of MPs. All self-enrichment models share the notion that a cluster will show MPs only if its mass is larger than a given threshold, i.e. only the most massive clusters should be able to retain the enriched ejecta of a first generation of stars (and accrete new material) to form a second generation. 
Hence, these models predict that clusters massive enough in the local universe should be undergoing multiple epochs of star-formation, in contradiction with observations \citep{bastian13,cabrera14}.
More recent versions have attempted to address this issue by invoking special conditions in the early Universe, at redshifts above $z_{\rm formation}\sim 2$ \citep{dercole16}. However, the discovery of chemical anomalies in NGC 1978 ($z_{\rm formation}=0.17$) leads to the conclusion that the onset of MPs cannot be limited to the high-$z$ Universe.

The data presented here tentatively suggest that MPs may be due to a stellar evolutionary effect not yet recognized in standard evolution models. 
This effect would need to only efficiently operate in stars within massive/dense stellar clusters.  
NGC 1978 is currently the youngest cluster for which detection of MPs have been reported. At this age ($\sim$2 Gyr), the sampled stellar mass in the lower part of the RGB is slightly lower than 1.5 \msun
while it is $\sim$1.55 \msun \, at 1.7 Gyr, i.e. the typical age of the young surveyed clusters where MPs are not found. One potentially important change in this very narrow mass range is that stars above $\sim$1.5 \msun \, do not typically possess strong magnetic fields, whereas stars below this mass do. 
This can also be related with stellar rotational properties, as stars with strong magnetic fields can be magnetically braked, leading to slow rotation rates, whereas stars with weak magnetic fields can remain rapid rotators (e.g. \citealt{cardini07}).  Additionally, we observe the eMSTO in the optical CMD of clusters younger than 1.7 Gyr (NGC 419, 1783,1806, 1846), while we do not observe such a feature in the $\sim$ 2 Gyr old NGC 1978 (see Fig. \ref{fig:S7}), where the mass of the lower RGB and MSTO stars has decreased below 1.5 \msun.
If the phenomenon is related to stellar rotation and/or magnetic fields, it is worth noting that the rates of stellar rotation are linked to the cluster mass, i.e. rotation can cause environmentally dependent stellar evolutionary effects\footnote{We note that the $\sim1.3$ Gyr lower mass open cluster, Trumpler 20, does not show signs of an extended main sequence turnoff.  However, it also does not appear to host rapidly rotating stars \citep{platais12}.}.

Another intriguing consequence of such an interpretation may be the expected presence of chemical anomalies in stars with masses below 1.5 \msun\ on the MS of the young clusters. Furthermore, one immediate implication, to account for the rare objects with M $< 10^5$ \msun\ seen in Fig. \ref{fig:ageM} that have ages older than 2 Gyr but no MPs, is that the global properties of a cluster (such as its mass, or initial angular momentum) can, under certain circumstances, influence the properties of its individual stars (such as the distribution of rotation speeds). Recent observations \citep{corsaro17} suggest that this is indeed plausible. However, this remains purely speculative at the moment. This is a completely unexplored direction for the onset of MPs, and further tests need to be carried out to confirm or refute such interpretations.

A major implication of our discovery of MPs in a young (2 Gyr) cluster, is that it lends support to the view that the ancient GCs and young massive clusters (YMCs) share a common formation process, as MPs have now been found in both classes of clusters.  While globular and young massive clusters overlap in many of their properties, such as mass, size and stellar density, many models for the formation of GCs have adopted special conditions only present in the early Universe \citep{trenti15}. One line of support for such a distinction between the ancient globulars and YMCs was that only globular clusters were thought to host MPs.  
The results presented here show that the formation of MPs continued at least down to a redshift of $z=0.17$, well past the peak epoch of GC formation in the Universe ($z=2-5$, \citealt{brodie06}).
Instead, our results support models that explain GC and YMC formation/evolution within a common framework \citep{kruijssen15}.

\section*{Acknowledgments}
We thank the referee, Raffaele Gratton, for his suggestions that helped improve the manuscript.
We, in particular F.N., N.B., I.P. and V. K.-P., gratefully acknowledge financial support for this project provided
by NASA through grant HST-GO-14069 for the Space Telescope Science Institute, which is operated by the Association
of Universities for Research in Astronomy, Inc., under NASA contract NAS526555. 
C.L. thanks the Swiss National Science Foundation for supporting this research through the Ambizione grant number PZ00P2\_168065. N.B. gratefully acknowledges financial support from the Royal Society (University Research Fellowship) and the European Research Council
(ERC-CoG-646928-Multi-Pop). We are grateful to Jay Anderson for sharing with us his ePSF software.
D.G. gratefully acknowledges support from the Chilean BASAL Centro de Excelencia en Astrof\'isica
y Tecnolog\'ias Afines (CATA) grant PFB-06/2007. 
Support for this work was provided by NASA through Hubble Fellowship grant \# HST-HF2-51387.001-A awarded by the Space Telescope Science Institute, which is operated by the Association of Universities for Research in Astronomy, Inc., for NASA, under contract NAS5-26555.

\bibliographystyle{mn2e}
\bibliography{hst_survey.bib}

\label{lastpage}
\end{document}